\documentstyle[aaspp4]{article}
\def \sun {$_{\scriptscriptstyle \odot}$}
\lefthead{FRYER}
\righthead{Rotating Supernovae}
\begin{document}
\begin{center} To be submitted to {\em The Astrophysical Journal}
\end{center}
\vspace{1.cm}
\title{Core-Collapse Simulations of Rotating Stars}
\author{Chris L. Fryer \& Alexander Heger}
\affil{Lick Observatory, University of California Observatories,
\\  Santa Cruz, CA 95064 \\ cfryer@ucolick.org, alex@ucolick.org}
\authoremail{cfryer@ucolick.org, alex@ucolick.org}

\begin{abstract}
We present the results from a series of two-dimensional core-collapse
simulations using a rotating progenitor star.  We find that the
convection in these simulations is less vigorous because a) rotation weakens
the core bounce which seeds the neutrino-driven convection and b) the
angular momentum profile in the rotating core stabilizes against
convection.  The limited convection leads to explosions which occur
later and are weaker than the explosions produced from the collapse of
non-rotating cores.  However, because the convection is constrained to
the polar regions, when the explosion occurs, it is stronger along the
polar axis.  This asymmetric explosion can explain the polarization
measurements of core-collapse supernovae.  These asymmetries also
provide a natural mechanism to mix the products of nucleosynthesis out
into the helium and hydrogen layers of the star.  We also discuss the
role the collapse of these rotating stars play on the generation of
magnetic fields and neutron star kicks.  Given a range of progenitor
rotation periods, we predict a range of supernova energies for the
same progenitor mass.  The critical mass for black hole formation also
depends upon the rotation speed of the progenitor.

\end{abstract}

\keywords{black hole physics - stars: rotation - supernova: general }

\section{Introduction}

The study of the collapse of {\it rotating} massive stars is nearly as
old as the study of core-collapse supernovae themselves.  Four years
after the first numerical simulations of neutrino powered
core-collapse supernovae (Colgate \& White 1966), LeBlanc \& Wilson
(1970) modeled the core-collapse of a rotating massive star.  Instead
of using neutrinos to convert gravitational energy into kinetic energy
and power the explosion (Colgate \& White 1966), LeBlanc \& Wilson
(1970) proposed that supernovae were powered by the conversion of
rotational energy using magnetic fields.  Although neutrino heating is
now considered to be the dominant power source driving supernovae, the
importance of rotation remains a matter of dispute.

Most of the recent work studying the collapse of rotating massive
stars has concentrated on the emission of gravitational waves
(M\"uller, R\'o\.zyczka, \& Hillebrandt 1980; Tohline, Schombert, \&
Boss 1980; Finn \& Evans 1990; M\"onchmeyer et al. 1991; Bonazzola \&
Marck 1993; Yamada \& Sato 1995; Zwerger \& M\"uller 1997; Rampp,
M\"uller, \& Ruffert 1998).  Some of the work, however, was devoted to
understanding the effect of rotation upon the supernova explosion
itself (M\"uller \& Hillebrandt 1981; Bodenheimer \& Woosley 1983;
Symbalisty 1984; M\"onchmeyer \& M\"uller 1989; Janka \& M\"onchmeyer
1989; Janka, Zwerger, \& M\"onchmeyer 1993; Yamada \& Sato 1994).
Currently, no consensus on the effects of rotation has been reached.
For instance, there is still disagreement on whether rotation
increases or decreases the explosion energy.  M\"onchmeyer \& M\"uller
(1989) found that rotation weakens the core bounce, ultimately
weakening the explosion.  On the other hand, the asymmetry in the 
neutrino emission caused by rotation may help seed convection, increasing 
the efficiency of neutrino heating and ultimately producing a more 
powerful explosion (Yamada \& Sato 1994).

The simplifying assumptions in all of these simulations make it 
difficult to resolve this disagreement.  For instance, none of the 
above simulations incorporated neutrino transport into their 
multi-dimensional hydrodynamic simulations.  Many of the simulations 
use extremely simplified equations of state which, as M\"uller \& 
Hillebrandt (1981) discovered, alters significantly the effect of 
rotation.  In addition, the pre-collapse cores used in these models 
were all created assuming no rotation, and the angular momentum 
was then added artificially prior to the collapse simulation.

Two major advances in supernova theory seek to resolve the question of
rotation effects on core-collapse.  First, in the past 5 years, codes
have been developed with the necessary equations of state and neutrino
physics to model the core collapse in two-dimensions from the initial
collapse all the way through explosion (Herant et al. 1994; Burrows,
Hayes, \& Fryxell 1995, Fryer 1999).  In addition to improvements on
the collapse codes, Heger (1998); Heger, Langer, \& Weaver (1999), 
has evolved rotating massive stars to
core-collapse using a prescription for the transport of angular
momentum, producing rotating core-collapse progenitors.  In this
paper, we present results of collapse simulations of these rotating
massive cores.

In \S 2, we outline the assumptions of the progenitor models and
discuss the specific progenitor we use in our simulations.  The core
collapse code is described in \S 3 and the basic effects of rotation
are outlined in \S 4.  We find that not only is the bounce of rotating
stars weaker, but the angular momentum in the star stabilizes the
core, constraining convection to the poles.  These two effects
decrease dramatically the efficiency at which convection is able to
convert neutrino energy into kinetic energy.  The net effect is to
delay the supernova explosion, producing larger compact remnants and
weaker explosions.  We present the explosion simulations for all our
models in \S 5.  The asymmetry in the convection produces asymmetric
explosions which may explain polarization measurements of supernovae
and may inject alpha-elements deep into the star's envelope.  We
conclude by reviewing the implications of these results on
core-collapse supernovae, black hole formation, and the collapsar
gamma-ray burst model.

\section{Progenitor}
All of the previous collapse simulations of rotating cores prescribed
an angular momentum profile onto a non-rotating progenitor.  For our
simulations, we use the 15\,M\sun rotating progenitor E15B of Heger 
et al. (1999).  At central hydrogen ignition, this 
model has a ``solar'' composition (Grevese \& Noels 1993)
and an equatorial rotation velocity of 200\,km\,s$^{-1}$ which 
is a typical value for these stars (e.g., Fukuda 1982).  For this model, 
redistribution of angular momentum and chemical mixing was formulated 
similar to the technique of Endal \& Sophia (1978) and used a 
parameterization of the different mixing efficiencies similar to that 
of Pinsonneault et al. (1989).  The rotationally induced
mixing processes include dynamical and secular shear instability,
Solberg-H{\o}iland instability, Goldreich-Schubert-Fricke instability,
and Eddington-Sweet circulation.  All these mixing processes, as well
as the convective instability, were assumed to lead to rigid rotation
on their corresponding time-scale (Fig. 1 shows the evolution of the 
angular momentum at various stages of the star's life).  Equipotential 
surfaces where assumed to be rigidly rotation and chemically homogeneous 
due to the barotropic instability and horizontal turbulence (Zahn 1975;
Chaboyer \& Zahn 1992; Zahn 1992).  Magnetic fields were not
considered, since their efficiency in redistributing angular momentum
(and mixing) inside stars is still very controversial (e.g., Spruit \&
Phinney 1998, Spruit 1998 vs. Livio \& Pringle 1998) and no reliable
prescriptions exist yet.  For more detail on the progenitor model and
the input physics we refer the reader to Heger (1998); Heger et al. (1999).

If magnetic fields are strong, the angular velocity between the core
and surface may couple during most evolutionary phase, leading to much
smaller core rotation rates than found by Heger et al. (1999).  
In the model for the evolution of an internal
stellar magnetic field by Spruit \& Phinney (1998), the core rotation
decouples from the surface only \emph{after} central carbon depletion.
The cores resulting from their calculation carry far too little
angular momentum to explain the observed rotation rates of young
pulsars (Marshall et al. 1998), and thus Spruit \& Phinney (1998) have 
to employ another mechanism to spin them up again.  Livio \& Pringle (1998) 
instead find that the coupling between core and envelope due to magnetic 
fields should be far less than assumed by Spruit \& Phinney (1998), 
allowing a natural explanation for spin rates of both young pulsars and 
white dwarfs.

With regard to the relevant physics in core-collapse and the
neutrino-driven supernova mechanism, the structure of our rotating
progenitor is not too different from that of Woosley \& Weaver (1995).
Figure 2 shows the density and temperature profiles just before
collapse for our rotating progenitor and model s15s7b from Woosley \&
Weaver (1995).  Note that beyond a radius of $\sim 3\times 10^8$\,cm,
the density of the non-rotating star is roughly a factor of 2 lower
than that of the non-rotating Woosley \& Weaver progenitor.  Pressure
equilibrium requires that the temperature at these radii also be
lower.  Burrows \& Goshy (1993) have shown that cores with lower
densities are more likely to explode because the ram pressure which
prevents the explosion is lower for these cores (see \S 4).  However,
the differences between these two stars are small enough that they do
not change the ultimate features in the core-collapse significantly
(\S 5).

The angular momentum distribution, on the other hand, does
dramatically effect the collapse.  The angular velocity profile of our
rotating progenitor is shown in Figure~3.  For comparison, we have
included the angular velocity profiles of many of the models used by
M\"onchmeyer \& M\"uller (1989).  Rather than use a rotating
progenitor, M\"onchmeyer \& M\"uller (1989) added these angular
velocities onto a non-rotating progenitor.  Their model A closely
resembles our rotating progenitor, but most of their models have
higher angular velocities, and the effects of rotation are stronger
for most of their simulations.  In addition, our angular velocity
distribution consists of a series of discreet shells which rotate at
constant angular velocity whereas the M\"onchmeyer \& M\"uller (1989)
models all assume a continuous rotation distribution.  These discreet
shells are the result of discreet convection regions in the progenitor
star.  The strong convection in the star causes the material in each
shell to reach a constant angular velocity (rigid body rotation).
However, shear viscosity does not equilibrate adjacent shells and, at
the boundaries of convective regions, the angular velocity can change
abruptly.  The evolution of the core collapse depends most sensitively
on the magnitude of the rotation speed and not its derivative and,
hence, the initial collapse of model A from M\"onchmeyer \& M\"uller
(1989) agrees well with our simulations (see \S 4).

\section{Numerical Techniques}
For our simulations, we use the smooth particle hydrodynamics (SPH)
core-collapse code originally described in Herant et al. (1994) and
Fryer et al. (1999).  This code models the core collapse in two
dimensions continuously from collapse through bounce and ultimately to
explosion.  The neutrino transport is mediated by a single energy
flux-limiter.  Beyond a critical radius, $\tau<0.1$, a simple
``light-bulb'' approximation for the neutrinos is invoked which
assumes that any material beyond that radius is bathed by an isotropic
flux equal to the neutrino flux escaping that radius.  We use all the
neutrino production and destruction processes discussed in Herant et
al. (1994) {\it except} electron scattering because our neutrino
transport algorithm overestimates its effect.  We implement gravity
assuming the mass distribution is spherical.  Assuming a spherical
mass distribution also allows us to easily implement general
relativity fully into the equations of motion using the technique
described by Van Riper (1979).  This treatment includes both the
relativistic modifications to the energy equation and time dilation.
We further red-shift the neutrino energies as the neutrinos climb out
of the core's potential well.

Our two-dimensional simulations assume cylindrical symmetry about 
the rotation axis, modeling a full 180$^\circ$ wedge with a reflective 
boundary around the symmetry axis.  The inner 0.01M\sun is also modeled 
as a reflective boundary which collapses along with the star as 
described in Herant et al. (1994).  Because the progenitor models 
did not account for centrifugal acceleration, we had to artificially 
add 0.04\,M\sun to the inner core to induce a collapse (hence, 
the inner core has an effective mass of 0.05\,M\sun).  We also ran a 
set of simulations with an additional 0.04\,M\sun to the core but 
the results did not change (Table 2).  For most of our simulations, 
we use $\sim 9000$ particles with a resolution in the convection 
parts of the star of roughly 2$^\circ$.  We have run one model with 
$\sim 32,000$ particles (twice the resolution) and the results do 
not change significantly (Table 2).

The initial angular momentum distribution of the progenitor (See \S 2;
Heger et al. 1999) is mapped onto the 2-dimensional grid of particles
assuming that the rotation axis is identical to the symmetry axis.
Each particle is given an initial specific angular momentum:
$j_i=(x_i/r_i)j(r)$ where $r_i$, $x_i$ are the radial distance and the
particle distance from the rotation axis and $j(r)$ is the angular
momentum given by E15B progenitor of Heger et al. (1999; Fig. 3).  We
have also set up models using $j_i=0.5(x_i/r_i)j(r)$ and $j_i=0.$ For
most of our simulations, we assume that the angular momentum is
conserved, which in our 2-dimensional SPH code is true simply by
holding $j_i$ constant for each particle.  The force on the particles,
which are rings in our rotationally symmetric simulation, is modified
by adding the centrifugal force: $\equiv j^2_i/x^3_i$.  By ranging the
initial angular momenta of the particles, we can calculate the effects
of rotation.

However, in our rotating models, the angular velocity of the 
matter ($v^\phi$) becomes so high that it may rotate 10-100 periods 
($P_i=2 \pi x_i/v^\phi_i$) during the course of the $\sim 1$\,s 
simulation.  In this case, it is likely that viscous forces will 
facilitate the transport of angular momentum.  To estimate the 
effects of these viscous forces, we have implemented an $\alpha-$disk
viscosity (Shakura \& Sunyaev 1973) which is suitable for the disk-like 
structures that develop as the core collapses.  We use the shear 
viscosity tensor from Tassoul (1978) and drop the terms involving 
a derivative in $\phi$ (since $\partial X/\partial \phi = 0$ in 
our rotational symmetry).  Using our SPH formalism, the transport 
of angular momentum is:
\begin{equation}
\frac{D j_i}{D t} = \sum_k \alpha_{\rm d} \bar{c}_s \bar{H}_{\rm p} 
\bar{x} \frac{\bar{m}}{m_i} 
\left [ \frac{3 \Delta (v^\phi/x)}{r_{ik} + \epsilon \bar{h}} +
\frac{\bar{x} \Delta (v^\phi/x)}{r_{ik}^2 + \epsilon \bar{h}^2} +
\frac{\bar{x} \Delta (v^\phi)}{r_{ik} + \epsilon \bar{h}} \right ]
\end{equation}
where $\alpha_{\rm d}=0.1-10^{-4}$ is the $\alpha-$disk parameter,
$\bar{c}_s$, $\bar{H}_{\rm p}$, $\bar{x}$, $\bar{m}$, and $\bar{h}$
are the mean values between neighbor particle $i$ and $k$ of,
respectively, the adiabatic sound speed, the disk scale height, the
distance from the rotation axis, the particle mass, and the SPH
smoothing length.  $\Delta (v^\phi/x)=(v^\phi_i/x_i-v^\phi_k/x_k)$,
$\Delta (v^\phi)=(v^\phi_i-v^\phi_k)$, $r_{ik}$ is the separation of
particles $i$ and $k$ and $\epsilon=0.1$.  The net energy lost by the
particles is then added to the particles equally as dissipation energy
which is equivalent to using the dissipation function given in Tassoul
(1978).  The summation occurs over all of the SPH neighbors of
particles for which $P_i<1$\,s.  This implementation conserves angular
momentum and energy and the $\alpha-$disk viscosity gives some insight
into the effects of viscous forces.

To follow the progression of the supernova explosion to late 
times ($\sim1\,$yr after core collapse), we convert from our 
2-dimensional results back into the one-dimensional 
stellar evolution code KEPLER (Weaver, Zimmermann, \& Woosley 1978).  
This is the same code that has been used to evolve the stars until core 
collapse (Heger 1998; Heger, Langer, \& Woosley 1999).  Unfortunately, 
KEPLER simulates the explosion using a ``piston'' rather than a simple 
injection of energy which matches the physics more reliably.  The piston 
moves inward at free-fall until then infall velocity reaches about 
1,000\,km\,s$^{-1}$ at which point it bounces and moves outward until it 
reaches a final position of $10^4$\,km.  The first part of the piston 
movement is taken form the SPH simulation by following the trajectory 
of a spherical mass shell.  This method leads to a systematic 
underestimate of the energy input, because any spherically determined 
mass shell has matter flowing through it and although the mean 
velocities of the exploding matter is matched by the spherically 
determined mass trajectories, total energy in the explosion can 
be much higher.  As the simulation progresses beyond the maximum time 
from our two-dimensional simulations, we continue the movement of the 
piston such that it resembles the free movement of a test particle in 
the gravitational field of a point mass equal to $\alpha$ times the mass 
interior to the piston which reaches zero velocity at the final location 
of the piston ($10^4$\,km; see above), following the prescription by 
Woosley \& Weaver (1995).  

The value of $\alpha$ is chosen such that it ``smoothly'' continues
the piston movements obtained from the SPH simulations.  This,
however, involves some ambiguity (about a factor of two in $\alpha$, we
estimate), since we have no measure to quantify the quality of this
fit.  The values of $\alpha$ we use are given in Table~3.  We carry
out these extended SN simulations for a rotating model (Model~1; see
Table~1) and a non-rotating model (Model~6) and for different
locations, in mass, of the piston (Table~3).  For Model~6 we switch to
the ``artificial'' piston after 1.9\,s after the onset of core
collapse, for Model~6 after 1.0\,s.  The contribution of this
artificial continuation of the SPH-derived pistons to the total work
of the piston is also given in Table~3: in most cases it is only small
and thus our choice of $\alpha$ should not significantly alter our
results.  We note that during the inward movement of the piston a work
of about $10^{51}$\,erg (for Model~1, 1.1\,M\sun piston) is done
against the piston by the infall and thus the work of the piston
counting form its minimum location (in radius) is higher by this
amount.  The piston energies given in Table~3 are the integrated work
energies from the beginning of the SPH simulation till $\sim1\,$yr
after the core collapse.

For the non-rotating Model~6 we also carry out a series of explosions
where we start the ``artificial'' piston at the minimum location of
the piston (as obtained from the SPH simulation) at a mass coordinate
of 1.1\,M\sun\ and for different values of $\alpha$.  In comparison to
the model of same mass cut but using the artificial piston only after
the end of the SPH data we get the same energies within less than
1\,\%.  The explosion energy turns out to be rather insensitive to the
value of $\alpha$.  Due to the above-mentioned underestimate of the
explosion energy in the 1D simulation, a more than six times higher
value of $\alpha$ ($\sqrt{6}$ times higher piston velocity) is
necessary to obtain the same explosion energy than in the SPH
simulation.  By using the results from this series of piston 
simulations, we can estimate not only the kinetic energies of 
our supernova explosions, but also the amount of fallback and 
$^{56}$Ni ejected (Table 2).

\section{Collapse and Convection}
The evolutionary scenario for the current paradigm of core-collapse
supernovae begins with the collapse of a massive star ($M_{\rm star}
\gtrsim 8\,$M\sun) which occurs when the densities and temperatures at
the core are sufficiently high to cause both the dissociation of the iron 
core (removing energy from the core) and the capture of electrons 
onto protons which further reduces the pressure.  This sudden decrease
in pressure causes the core to collapse nearly at free-fall, and the
collapse stops only when nuclear densities are reached and nucleon
degeneracy pressure once again stabilizes the core and drives a bounce
shock back out through the core.  The bounce shock stalls at roughly
100-300\,km as neutrino emission and iron dissociation sap its energy.
It leaves behind an entropy profile which is unstable to convection,
and sets up a convective region from $\sim$50\,km out to $\sim$300\,km
(Fig.~4) capped by the accretion shock of infalling material.  The
convective region must overcome this cap to launch a supernova
explosion.

Neutrino heating deposits considerable energy into the bottom layers
of the convective region.  If this material were not allowed to
convect (which is the case for most of the 1-dimensional simulations),
it would then re-emit this energy via neutrinos producing a steady
state with no net energy gain.  In the meantime, the increase in
pressure as more material piles up at the accretion shock and the
decrease in neutrino luminosity as the proto-neutron star cools make
it increasingly difficult to launch an explosion.  In the
``delayed-neutrino'' supernova mechanism (Wilson \& Mayle 1988;
Miller, Wilson, \& Mayle 1993; Herant et al. 1994; Burrows, Hayes, \&
Fryxell 1995; Janka \& M\"uller 1996), convection aids the explosion
in two ways: a) as the lower layers of the convective region are
heated, that material rises and cools adiabatically and converts the
energy from neutrino deposition into kinetic and potential energy
rather than re-radiating it as neutrinos, and b) the material does not
simply pile at the shock but instead convects down to the surface of
the proto-neutron star where it either accretes onto the proto-neutron
star providing additional neutrino emission or is heated and rises
back up.  Thus, convection both increases the efficiency at which
neutrino energy is deposited into the convective region and reduces
the energy required to launch an explosion by reducing the pressure at
the accretion shock.  It appears that nature has conspired to make
core-collapse supernovae straddle the line between success and
failure, where convection, which for years was thought to be a mere
detail, plays a crucial role in the explosion.  It is not surprising
that the outcome of the core collapse of massive stars depends upon
numerical approximations of the physics, for example, the algorithm
for neutrino transport (Messer et al. 1998; Mezzacappa et al. 1998a).
It is also not surprising that changes in the progenitor, i.e.  the
subject of this paper, rotation, also can lead to different outcomes.

For a rotating core-collapse, this basic evolutionary history remains
the same, but many aspects of the collapse, bounce, convection, and
explosion phases change.  To determine these variations, we compare a
rotating star (E15B from Heger et al. 1999) with
its non-rotating counterpart (Unless otherwise stated, all
quantitative results and figures compare Models 1 and 6).  First,
angular momentum slows the collapse and delays the bounce
(M\"onchmeyer \& M\"uller 1989; Janka \& M\"onchmeyer 1989).  For our
rotating model, the bounce occurred 200\,ms into the simulation, a full
50\,ms longer than the non-rotating counterpart.  For very high
angular momenta, the central density at bounce can be lower than a
non-rotating star by over an order of magnitude (M\"onchmeyer \&
M\"uller 1989; Janka \& M\"onchmeyer 1989), but for the more modest
angular momenta from Heger et al. (1999), the central density at bounce drops
from $4 \times 10^{14}$\,g\,cm$^{-3}$ for a non-rotating star to $3
\times 10^{14}$\,g\,cm$^{-3}$ for the rotating case.  This lower
critical bounce density occurs because the centrifugal force begins to
provide significant support (it increases from 2\% to 10\% that of the
gravitational force over the course of the simulation).  The resulting
bounce shock is weaker and, along the poles, stalls at lower radii
than the non-rotating case.  Along the equator where the effective
gravity is less, the accretion shock moves much further, and the star
quickly loses its spherical symmetry (Fig. 5).  60\,ms after the
bounce, the accretion shock of the rotating model is at $\sim
160,300$\,km respectively for the poles,equator.  At that time, the
spherically symmetric accretion shock of the non-rotating star is at
$\sim 300$\,km.

The position of the accretion shock, initially set by the position at
which the bounce stalls, is important for the success or failure of
the convection-driven supernova mechanism because it determines the
ram pressure that the convective region must overcome to drive an
explosion.  Even more important, however, is the entropy profile that
is left behind after the bounce fails.  Because the bounce is weaker
for rotating stars, the angle-averaged entropy decreases with
increasing angular momentum (Fig. 6).  The entropy in the strong shock
limit is $\propto v_{\rm shock}^{1.5}$, and along the equator, where
the material is collapsing less quickly, the entropy is much lower
(Fig. 7).  The steep entropy gradient in supernovae is what seeds the
convection and the shallower gradients in the collapse of rotating
stars lead to much weaker convection and, as we shall see in \S 5,
weaker explosions.

Rotation further weakens the explosion because the angular momentum
profile is stable to convection and constrains convection to the polar
regions.  This can be understood physically by estimating the change
of force on a blob of material as it is slightly raised from a
position $r$ with density ($\rho$), angular momentum ($j$) and
pressure ($P$) to a position $r+\Delta r$ with its corresponding
density, angular momentum, and pressure ($\rho+\Delta \rho, j+\Delta
j, P+\Delta P$).  If the blob is allowed to reach pressure equilibrium
in its new surroundings, the acceleration that it feels without
rotation is simply: $\Delta a=\Delta a_{\rm buo}=g(1-(\rho+\Delta
\rho)/ \rho_{\rm blob})$ where $g$ is the gravitational acceleration and
$\rho_{\rm blob}$ is the density of the blob of material after it has
expanded/contracted to reach pressure equilibrium.  If the change in
acceleration is positive, the blob of material will continue to rise
and that region is convectively unstable.  This gives us the standard
Schwarzschild/Ledoux instability criteria (Ledoux 1947): a region is
unstable if $(\partial \rho/\partial r)< (\partial \rho/\partial
r)_{\rm adiabat}$, or, for constant composition,  
$\partial S/\partial r > 0$.  If there is an
angular momentum gradient, however, the net force on the blob becomes:
\begin{equation}
\Delta a = g \left ( 1-\frac{\rho+\Delta \rho}{\rho_{\rm blob}} \right )+
\frac{j^2_{\rm blob}-(j+\Delta j)^2}{(r+\Delta r)^3}
\end{equation}
where $j_{\rm blob}=j$.  The corresponding Solberg-H{\o}iland
instability criterion is (Endal \& Sofia 1978):
\begin{equation}
\frac{g}{\rho} \left [ \left ( \frac{d \rho}{d r} \right
)_{\rm adiabat} - \frac{d \rho}{d r} \right ] >
\frac{1}{r^3} \frac{d j^2}{d r}
\end{equation}
If the angular momentum increases with increasing radius as it does
for our core collapse models (Fig. 8), then the entropy gradient must
overcome the angular momentum gradient to drive convection.  In our
simulations, the high entropy bubbles are unable to rise through the
large angular momentum gradient and the convection is constrained to
the polar region.  The overwhelming effect of rotation on supernova
models is this constraint on the convection and it causes weaker,
asymmetric explosions.

\section{Explosions And Compact Remnants}

Because the convection is limited in rotating models, the convective
region has less energy and is unable to overcome the pressure at the
accretion shock as early as the non-rotating models which explode
within 200\,ms of bounce.  However, at later times (500\,ms after
bounce) when the density of the infalling material decreases, the
rotating models are able to launch an explosion, but the explosion
occurs later and the proto-neutron star accretes slightly more mass
(Fig.~9, Table~2).  The final compact remnant mass, however, depends
upon the amount of material that falls back onto the proto-neutron
star.  By mapping our explosions back into one-dimension, we can
estimate the amount of fallback and the observed explosion energy
(Table~2).  The masses in Table~2 are baryonic masses and the
gravitational masses, which can be compared to observations, are
typically $\sim$10\,\% lower.  Mass estimates from observations of
close binaries predict gravitational masses for stars without hydrogen
envelopes to be $\sim$1.4\,M\sun.  Even
including fallback, our non-rotating models produce remnants with
masses which are much lower than those observed.  Although the remnant
masses of rotating models match observations much better, the
explosion energies are too low to explain all the observations.  Due
to the uncertainties in other aspects of the supernova simulation
(e.g., neutrino physics), we can not claim that rotating stars are
required to make close compact binaries or that non-rotating stars are
required to explain most supernova observations.  But we can be
reasonably sure of the trends: rotating stars produce weaker
explosions and more massive remnants.  As a further consequence, the
rotating stars eject notably less $^{56}$Ni (Table~2), up to one half
for some of our explosions.

Not only are the explosions weaker for rotating stars, but since the
convection is constrained to the poles, the explosion is stronger
along the poles (Figures 10, 11).  At the end of the simulation, $\sim
1.5$\,s after bounce, the shock radius is 1.4 times further out along
the pole than along the equator for our rotating star (Table~2).  The
mean velocity of the shocked material being ejected at the end of the
simulation (2/3 of the explosion energy is still thermal) is
13,000\,km\,s$^{-1}$ along the pole and only 5,100\,km\,s$^{-1}$ along
the equator!  The maximum velocity along the pole can be as high as
18,000\,km\,s$^{-1}$.  These results must be taken with some degree of
caution, since the axis of symmetry lies along the polar axis, and
some of this asymmetry could be due to numerical artifacts.  However,
if we compare these results with a non-rotating star, we are
encouraged that this asymmetry is real.  In the case of a non-rotating
star, the shock radius is only 4\,\% further out along the poles, and
the mean polar, equatorial velocities are 8,000\,km\,s$^{-1}$,
7,700\,km\,s$^{-1}$ respectively (maximum velocities are as high as
20,000\,km\,s$^{-1}$).  The variations from pole to equator in the 
non-rotating simulations are likely to be caused by the position of 
the rising bubbles and down-flows during the explosion.  Even if these 
variations were all due to numerical artifacts (it could be due to convection 
plumes and hence, physical), it appears that numerical artifacts can account 
for only 10\,\% variations in the velocities and the large asymmetry in the
rotating simulations are likely to be real.  By constraining the
convection to the poles, rotating models produce more than a factor of
2 asymmetry in the explosion.

Neutrino luminosities for rotating and non-rotating stars are plotted
in figure 12.  The non-rotating core has a much larger $\mu$ and
$\tau$ ($\nu_{\rm x}$) neutrino luminosity, especially just after
bounce.  This is because the non-rotating core compresses more and, at
the $\mu$ and $\tau$ neutrinosphere, the temperature is over a factor
of 1.5 higher than that of the rotating core.  The large dependence 
of neutrino emission on temperature (the luminosity from pair 
annihilation $\propto T^9$), causes this small change in temperature 
to have large effects on the neutrino luminosity.  Also, bear in mind that
beyond 0.2\,s after bounce, the convective regions in the star look
dramatically different.  The non-rotating star has already launched an
explosion, leaving a hot, bare proto-neutron star which continues to
cool by emitting neutrinos.  The rotating core is still convecting,
and its neutrinosphere is further out.  These differences are more
easily seen in a plot of neutrino energies (Fig. 13).  We infer no net
asymmetry in the neutrino luminosity.  But because of our coarse
resolution, and the rapid time variation of the neutrino luminosity as
bubbles rise above the neutrinosphere, asymmetries of $\sim 20-30$\%
could be hidden in our data and we can neither confirm nor rule out
the asymmetric neutrino emission seen by Janka \& M\"onchmeyer (1989).

Rotating progenitors produce rotating compact remnants.  Table 2 lists
the spin period of the compact remnant assuming solid body rotation
both at the end of the simulation before the neutron star has cooled
and after it has cooled to a 10\,km neutron star.  Even for Model 3
(Fig. 8, Table 2), which includes the transport of angular momentum
using an $\alpha$-disk prescription, at the end of the simulation, the 
angular momentum in the proto-neutron star is sufficient to produce
a ms pulsar.  This is because, at the end of our SPH simulations, the 
proto-neutron star is still large, and it is not rotating rapidly.  
Hence, viscous momentum transport does not have an opportunity to 
remove the angular momentum from the core.  As the neutron star contracts 
and the spin increases, the neutron star will lose angular momentum through 
a wind.  Hence, the spin period of the cooled neutron star should be seen as a
lower limit.  In addition, convection in the cooling proto-neutron
star (Burrows, Mazurek, \& Lattimer 1981; Keil, Janka, \& M\"uller
1996) will increase the efficiency of the viscous transport of angular
momentum.  Even in the absence of any wind, gravitational radiation
limits the spin period to $\sim$3\,ms (Lindblom, Mendell, \& Owen
1999).  Fallback may be able to spin up the neutron star, but will not
be able to exceed the limit from gravitational radiation because the
accreting material heats the neutron star to temperatures where
gravitational radiation will effectively remove the angular momentum
(Fryer, Colgate, \& Pinto 1999).

A common problem to all core-collapse simulations which produce 
explosions is the large amount of neutron rich ejecta.  Although 
the amount of material ejected with extremely low
electron fraction ($Y_{\rm e} < 0.4$) decreases from $1.2 \times
10^{-3}$\,M\sun for the non-rotating model to $1.1 \times
10^{-5}$\,M\sun for the rotating case, but due to the delayed
explosion which gives more time for neutrino emission to deleptonize
the ejecta, the amount of mildly neutron rich material ($Y_{\rm e} <
0.49$) actually increases: 0.36, 0.60\,M\sun respectively for the
non-rotating, rotating models.  The material which falls back onto the
neutron star will be mostly comprised of this neutron rich material.
However, even assuming only neutron rich material falls back onto the
neutron star, $>0.1$\,M\sun\ of neutron rich material is ejected, several
orders of magnitude greater than the $10^{-2}-10^{-3}$\,M\sun
constraint from nucleosynthesis (Trimble 1991).  Clearly, the delayed
explosion caused by rotating does not solve the nucleosynthesis
problem of the current exploding core-collapse models.  

\section{Implications}
Rotation limits the convection in core-collapse supernovae, both by
weakening the shock and by constraining the convection to the polar
regions.  Because of these constraints, the convective region takes
longer to overcome the accretion shock and the explosion occurs at
later times and is less energetic.  The resultant compact remnants of
the collapse of rotating stars are more massive.  For higher mass
progenitors, the density of the infalling material, and hence the
shock pressure, increases (Burrows \& Goshy 1993).  Because of their
limited convection, rotating cores are less likely to overcome the
accretion shock, and the critical mass limits for black hole formation
(both from fallback and direct collapse) are likely to be lower than
those predicted for non-rotating stars (see Fryer 1999).  Hence, the
same rotation required to power a collapsar (MacFadyen \& Woosley
1999) also causes the supernova to fail and the formation of a black
hole.  We expect a range of progenitor mass limits for black hole 
formation depending on the core rotation.  This may help to explain the
observed range of black hole systems (Ergma \& van den Heuvel 1998).
We stress that the uncertainties in the numerical implementation of
the physics (e.g., neutrino transport, neutrino cross-sections,
equation of state, symmetric gravity) limit our ability to make
quantitative estimates, and the numbers should be regarded merely as
``best-estimates''.  The trends (e.g., weaker explosions, lower black
hole formation limits) are more secure.

If angular momentum were conserved, the rotation periods of neutron 
stars formed from our progenitor model would be much faster than that 
expected in nascent neutron stars.  However, winds and gravitational 
radiation will remove much of the angular momentum.  If anything, 
it will be difficult to explain any rapidly spinning nascent neutron 
stars.

Not only are the explosions of rotating core-collapses weaker, but
they are highly asymmetric (roughly a factor of 2 in the mean
velocities from pole to equator).  Polarization measurements of
core-collapse supernova suggest that most supernovae are polarized
(Wang et al. 1996; Wang \& Wheeler 1998).  In modeling Supernova
1993J, H\"oflich et al. (1995) required explosions with asymmetries of
a factor of 2, roughly the same asymmetry we see in our simulations,
and hence, rotation provides a simple explanation of polarization
measurements.  These asymmetric explosions also provide a simple
explanation for the extended mixing in supernovae.  Observations of
Supernova 1987A in the X-rays (Dotani et al. 1987, Sunyaev et
al. 1987), $\gamma$-rays (Matz et al. 1988), and even in the infra-red
(Spyromilio, Meikle, \& Allen 1990) all push toward extended mixing of
iron-peak elements.  The matter ejected along the poles, with their
much higher velocities, will mix deeper into the star.  To
quantitatively determine the amount of mixing and polarization
produced by the asymmetric explosions of our rotating cores,
multi-dimensional simulations must be run out to late times.

We do not model magnetic fields in our simulations, but we can place
some constraints on their importance in the supernova explosion.
Thompson \& Duncan (1993) have argued against primordial magnetic
fields for neutron stars and, instead, suggest that strong neutron
star magnetic fields are produced in dynamos during the supernova
explosion itself.  They propose a high Rossby number ($Ro=P_{\rm
rot}/\tau_{\rm con}$ where $P_{\rm rot}$ is the rotation period and
$\tau_{\rm con}$ is the convective turnover timescale) dynamo, which
takes advantage of the fast convection velocities, and hence high
rossby numbers, of core-collapse supernovae.  However, this dynamo
requires many convective turnovers ($\tau_{\rm con} \lesssim 10^{-3}\,
T_{\rm exp}$, where $T_{\rm exp}$ is the explosion time) to
significantly magnify the magnetic fields.  Unfortunately, for our
rotating supernovae, with typical convection length scales and
velocities of $\sim 400$\,km, $\sim 4000$\,km\,s$^{-1}$ respectively,
the explosion is launched before convection can produce a strong
magnetic field.  In addition, until the proto-neutron star cools and
contracts, the rotation period of the disk-like structure is also slow
(Table 2), and a rotation driven dynamo will not occur until after the
supernova explosion.

As the proto-neutron star cools and shrinks, however, the Thompson \&
Duncan (1993) dynamo, driven by Ledoux convection (see Burrows,
Mazurek, \& Lattimer 1981; Keil, Janka, \& M\"uller 1996) is more
promising.  The strength of this convection is still a matter of
debate (Mezzacappa et al. 1998b), but if it is as strong as Keil,
Janka, \& M\"uller (1996) predict, magnetic fields in excess of
$10^{15}$G are obtainable at late times using the convection-driven
dynamo of Thompson \& Duncan (1993).  These magnetic fields will
enhance the angular momentum lost in the proto-neutron star wind.  But
in all of our simulations, this occurs after the launch of the
explosion, and the magnetic field would not affect the supernova
explosion.

Lastly, and most speculatively, we come to the topic of neutron star
kicks.  There is a growing set of evidence that neutron stars are born
with high space velocities with a mean magnitude $\sim
450$\,km\,s$^{-1}$ (see Fryer \& Kalogera 1998 for a review).  Rotation
breaks one symmetry, and it is tempting to speculate that it is then
easy to break an additional symmetry causing a net momentum in the
ejecta which is non-zero.  To balance out the momentum, the neutron
star must then have gained some momentum.  Indeed, the net momentum of
the ejecta along the poles in our simulation is
$6\times10^{39}$\,g\,cm\,s$^{-1}$.  This corresponds to a kick velocity
of only $30$\,km\,s$^{-1}$, over an order of magnitude too small to
explain the observed pulsar velocity distribution.  And this momentum
may be entirely a numerical artifact.  Asymmetric neutrino emission
can also cause neutron star kicks, but we detect no significant
neutrino asymmetry.  In any event, calculating quantitative results on
neutron star kicks requires 3-dimensional simulations where a fixed 
inner boundary is not necessary and the proto-neutron star is allowed 
to move.

\acknowledgements This research has been supported by NASA (NAG5-2843,
MIT SC A292701, and NAG5-8128), the NSF (AST-97-31569), the US DOE
ASCI Program (W-7405-ENG-48), and the Alexander von Humboldt-Stiftung
(FLF-1065004).  It is a pleasure to thank Stan Woosley, Andrew
MacFadyen and Norbert Langer for encouragement and advice.

\begin{deluxetable}{lcc}
\tablewidth{18pc}
\tablecaption{Rotating Models}
\tablehead{ \colhead{Model} & \colhead{Angular}  
& \colhead{$M_{\rm core}$} \\
\colhead{Number} & \colhead{Momentum}  
& \colhead{(M\sun)}} 

\startdata
% should use \\ instead of \nl

1 & $J_{\rm Heger}(r)$ &0.05 \\
2 & $J_{\rm Heger}(r)$ &0.09 \\
3\tablenotemark{a} & $J_{\rm Heger}(r)$ & 0.05 \\
4\tablenotemark{b} & $J_{\rm Heger}(r)$ & 0.05 \\
5 & $0.5 \times J_{\rm Heger}(r)$ & 0.05 \\
6 & 0. & 0.05 \\
7 & 0. & 0.09 \\
8\tablenotemark{c} & 0. & 0.01 \\

\tablenotetext{a}{High resolution run $\sim$32,000 particles}
\tablenotetext{b}{This model includes an $\alpha$-disk prescription 
for the artificial viscosity with $\alpha_{\rm d}$ set to 0.1 (\S 3).}
\tablenotetext{c}{This simulation uses the progenitor model s15s7b from 
Woosley \& Weaver (1995).  All other simulations use the progenitor 
described in \S 2 (Heger 1998).}

\enddata
\end{deluxetable}
\clearpage

\begin{deluxetable}{lcccccccc}
\tablewidth{40pc}
\tablecaption{Rotating Models}
\tablehead{ \colhead{Model} & \colhead{$T_{\rm exp}$\tablenotemark{a}}
& \colhead{$E_{\rm exp}$\tablenotemark{b}} 
& \colhead{M$_{\rm rem}$\tablenotemark{c}} 
& \multicolumn{2}{c}{Explosion Asymmetry} & \colhead{Rotation\tablenotemark{d}} 
& \multicolumn{2}{c}{Ejecta (M\sun)\tablenotemark{e}} \\
\colhead{$\#$} & \colhead{s}
& \colhead{$10^{51}$\,ergs}& \colhead{M\sun} 
& \colhead{$r_{\rm shock}^{\rm pole}/r_{\rm shock}^{\rm equ}$}
& \colhead{$V_{\rm pole},V_{\rm equ}$} 
& \colhead{ms} & \colhead{$Y_e<0.49$} & \colhead{$^{56}$Ni}}

\startdata

1 & 0.5 & 0.95,0.5 & 1.2,1.5 & 1.4 & 13,5.1 & 260,1.8 & 0.60,0.30 & 0.15 \\
2 & 0.5 & 1.2,0.5 & 1.2,1.5 & 1.3 & 12,6.1 & 140,1.7 & 0.56,0.25 & 0.2 \\
3\tablenotemark{f}  & 0.5 & - & - & - & - & - & - & \\
4 & 0.5 & 1.5,0.7 & 1.1,1.3 & 1.3 & 12,5.5 & 70,1.9 & 0.45,0.25 & 0.2 \\
5 & 0.4 & 1.9,1.3 & 1.2,1.3 & 1.2 & 12,9.0 & 260,1.9 & 0.50,0.40 & 0.3 \\
6 & 0.2 & 2.1,1.5 & 1.1,1.2 & 1.04 & 8.0,7.7 & - & 0.36,0.25 & 0.3 \\
7 & 0.2 & 2.2,1.6 & 1.1,1.2 & 1.06 & 9.7,9.2 & - & 0.35,0.25 & 0.3 \\
8 & 0.2 & 2.5,1.9 & 1.1,1.2 & - & - & -  & 0.23,0.1 & 0.3 \\

\tablenotetext{a}{The explosion time is the time from bounce until 
the shock has reached 2000\,km.}
\tablenotetext{b}{The explosion energy is given as both the 
net change in energy for the ejected material (see Fryer 1999) and the kinetic 
energy at infinity.}  
\tablenotetext{c}{The two values for remnant mass are before and after 
fallback.}
\tablenotetext{d}{The rotation period of the neutron star at 
the end of the SPH simulation (when the star is still rather 
large) and after the neutron star has cooled assuming no angular 
momentum is lost.}
\tablenotetext{e}{The first column gives the amount of neutron 
rich material ejected both before and after fallback.  The second 
column gives the total nickel mass ejected.}
\tablenotetext{f}{We have only run the high-resolution simulation 
to explosion (first $\sim 0.5$\,s) and can not reliably estimate 
the explosion energy or amount of fallback.  However, the explosion 
time and behavior agrees with the lower resolution runs.}

\enddata
\end{deluxetable}
\clearpage

\begin{deluxetable}{lcccccccc}
\tablecaption{1D explosion simulations}
\tablehead{ 
\colhead{Model} & 
\colhead{piston} &
\colhead{piston} &
\colhead{piston} &
\colhead{art. pist.} &
\colhead{explosion} &
\colhead{fall} &
\colhead{remnant} & 
\colhead{$^{56}$Ni}
 \\
\colhead{\#} & 
\colhead{location} &
\colhead{alpha} &
\colhead{energy\tablenotemark{a}} &
\colhead{energy\tablenotemark{a}} &
\colhead{energy\tablenotemark{b}} &
\colhead{back} &
\colhead{mass} & 
\colhead{ejected}
 \\
\colhead{} & 
\colhead{M\sun} &
\colhead{(art. part)} &
\colhead{$10^{51}$\,ergs} & 
\colhead{$10^{51}$\,ergs} &
\colhead{M\sun} &
\colhead{M\sun} &
\colhead{M\sun} 
}
\startdata
1\tablenotemark{c} & 1.10    & 0.0017 & 1.30 & 0.30 & 0.52 & 0.30 & 1.40 & 0.18 \\ 
1                  & 1.145   & 0.004  & 1.03 & 0.15 & 0.48 & 0.26 & 1.40 & 0.16 \\
1                  & 1.19    & 0.01   & 0.89 & 0.08 & 0.45 & 0.24 & 1.43 & 0.13 \\
1                  & 1.24    & 0.017  & 0.76 & 0.05 & 0.40 & 0.23 & 1.47 & 0.10 \\
1                  & 1.29    & 0.035  & 0.63 & 0.04 & 0.34 & 0.21 & 1.50 & 0.06 \\
\tableline  	     	      
6                  & 1.05    & 0.002  & 1.43 & 0.23 & 0.67 & 0.24 & 1.29 & 0.21 \\
6                  & 1.10    & 0.03   & 1.35 & 0.08 & 0.73 & 0.16 & 1.26 & 0.24 \\
6                  & 1.145   & 0.05   & 1.27 & 0.05 & 0.72 & 0.14 & 1.28 & 0.22 \\
6                  & 1.19    & 0.07   & 1.09 & 0.05 & 0.64 & 0.15 & 1.34 & 0.18 \\
6                  & 1.24    & 0.1    & 0.97 & 0.03 & 0.60 & 0.15 & 1.39 & 0.14 \\
6                  & 1.29    & 0.12   & 0.80 & 0.03 & 0.49 & 0.16 & 1.45 & 0.07 \\
\tableline  	     	      
6\tablenotemark{d} & 1.10    & 0.03   & 1.34 & 2.31 & 0.73 & 0.15 & 1.25 & 0.24 \\
6\tablenotemark{d} & 1.10    & 0.04   & 1.41 & 2.38 & 0.80 & 0.15 & 1.25 & 0.24 \\
6\tablenotemark{d} & 1.10    & 0.10   & 1.74 & 2.71 & 1.13 & 0.13 & 1.23 & 0.27 \\
6\tablenotemark{d} & 1.10    & 0.17   & 2.05 & 3.02 & 1.44 & 0.11 & 1.21 & 0.29 \\
6\tablenotemark{d} & 1.10    & 0.19   & 2.13 & 3.10 & 1.52 & 0.10 & 1.20 & 0.29 \\
\enddata
\tablenotetext{a}{Integrated work of the piston}
\tablenotetext{b}{Kinetic energy (of the ejecta) at infinity ($\sim
1\,$yr after explosion)}

\tablenotetext{c}{at 18.5\,s after core collapse the innermost 0.2
M\sun above the piston were removed from the grid and the new piston
radius (10$^9$\,cm) kept constant.  The deviation in the piston energy
due to this modification is $<1\,\%$.}

\tablenotetext{d}{artificial piston is started when Lagrangian mass
element reaches its minimum location}

\end{deluxetable}
\clearpage

\begin{figure}
\plotfiddle{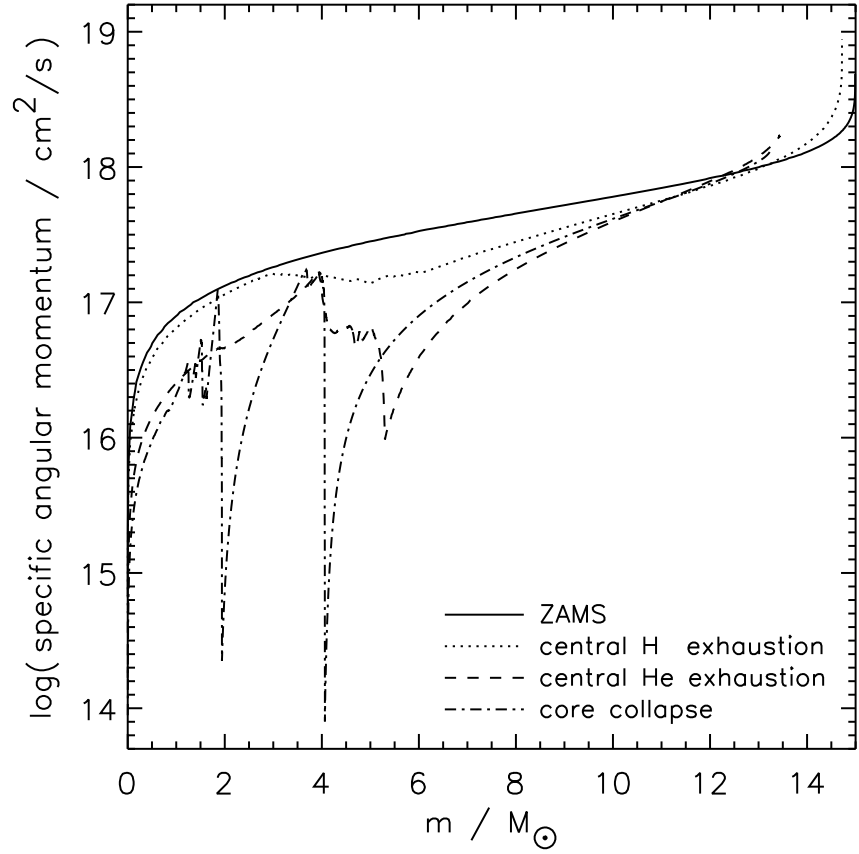}{7in}{0}{150}{150}{-480}{-300}
\caption{Specific angular momentum of the progenitor Model E15B at
different evolutionary stages as a function of the interior mass
coordinate.  The profiles at core hydrogen ignition (ZAMS;
solid line), central hydrogen depletion (dotted line), central helium
depletion (dashed line), and at onset of core collapses (dash-dotted
line) are given.  See also Heger (1998); Heger, Langer \& Woosley (1999).}
\end{figure}
\clearpage

\begin{figure}
\plotfiddle{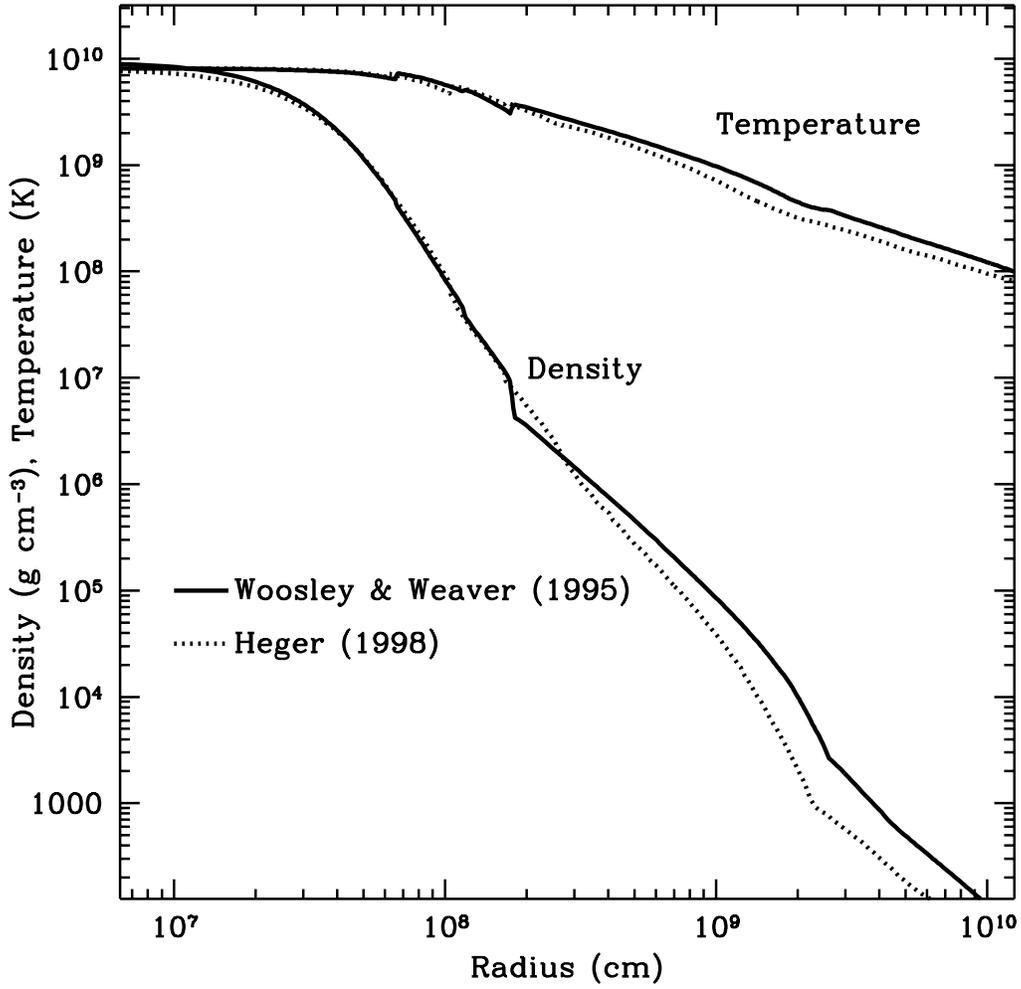}{7in}{0}{70}{70}{-210}{-10}
\caption{Density and Temperature profiles for our 15\,M\sun 
rotating progenitor (Heger 1998) and the Woosley \& Weaver 
(1995) progenitor s15s7b.  Beyond a radius of 
$3\times10^8\,$g\,cm$^{-3}$, the density of the rotating 
progenitor drops faster than s15s7b.  Pressure equilibrium 
requires that the temperature decrease faster as well.  
Although this lower density makes it easier to drive a supernova 
explosion, its effect is small compared to the effects of rotation.}
\end{figure}
\clearpage

\begin{figure}
\plotfiddle{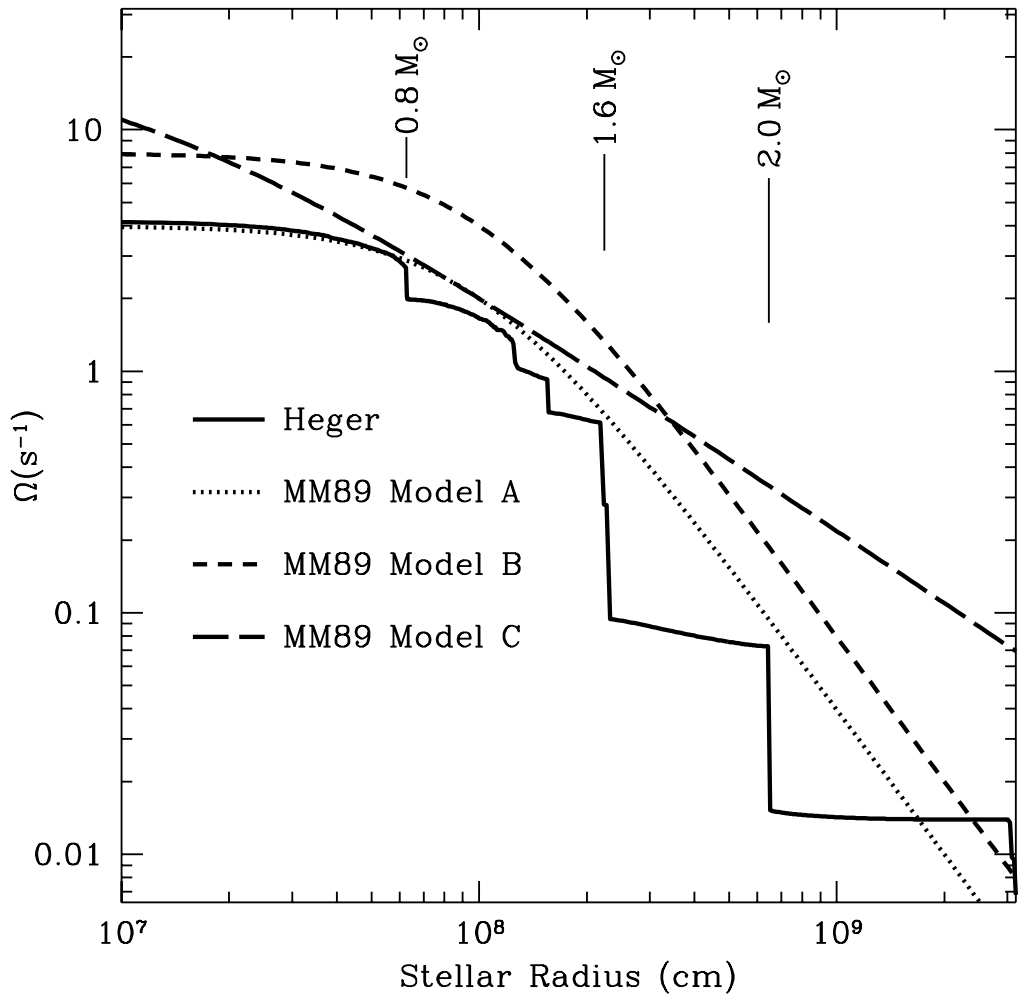}{7in}{0}{70}{70}{-220}{-10}
\caption{The angular velocity profile for our rotating model 
as well as the prescriptions used in the models of M\"onchmeyer 
\& M\"uller (1989).  Except for model A, the models of 
M\"onchmeyer \& M\"uller (1989) are rotating more rapidly than 
our models and the effects of rotation on our collapse simulations 
are much less than most of their simulations.  Also note that 
the angular velocity distribution is a series of step functions.  
This is because each convection region tends to equilibrate 
the angular velocity and only at convective boundaries does 
the angular velocity change significantly.}
\end{figure}
\clearpage

\begin{figure}
\plotfiddle{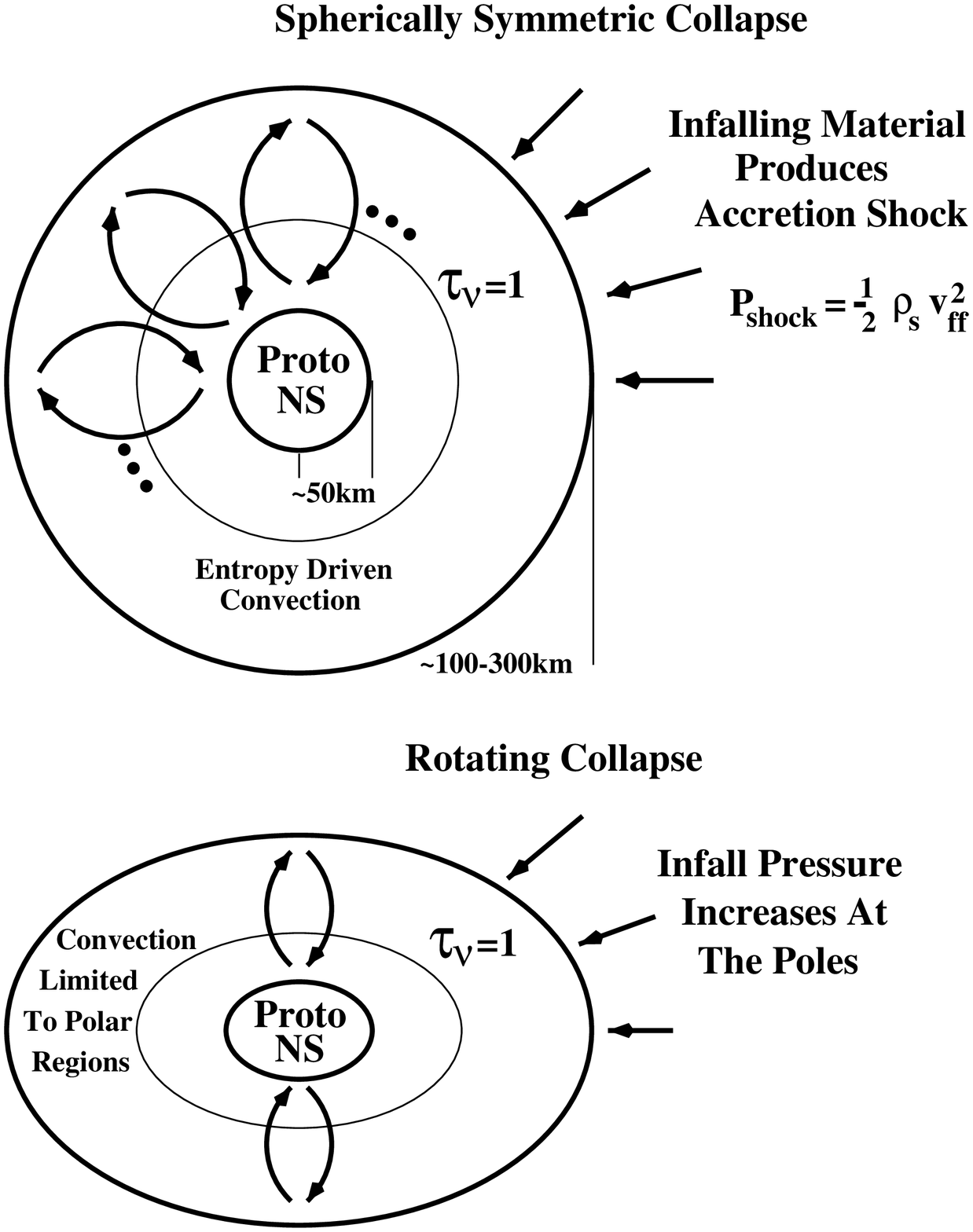}{7in}{0}{70}{70}{-210}{-10}
\caption{Entropy-driven convection has been found to enhance 
the energy deposition within the accretion shock.  As the 
density of the infalling matter decreases, the convective 
region is able to push the shock outward, launching a supernova 
explosion.  The proto-neutron star may also convect 
(Burrows, Mazurek, \& Lattimer 1981; Keil, Janka, \& M\"uller 1996), 
increasing the neutrino flux at its surface and the neutrino 
heating in the convective region.  The angular momentum profile 
in rotating core-collapses stabilizes the convection along the 
equator.}
\end{figure}
\clearpage

\begin{figure}
\plotfiddle{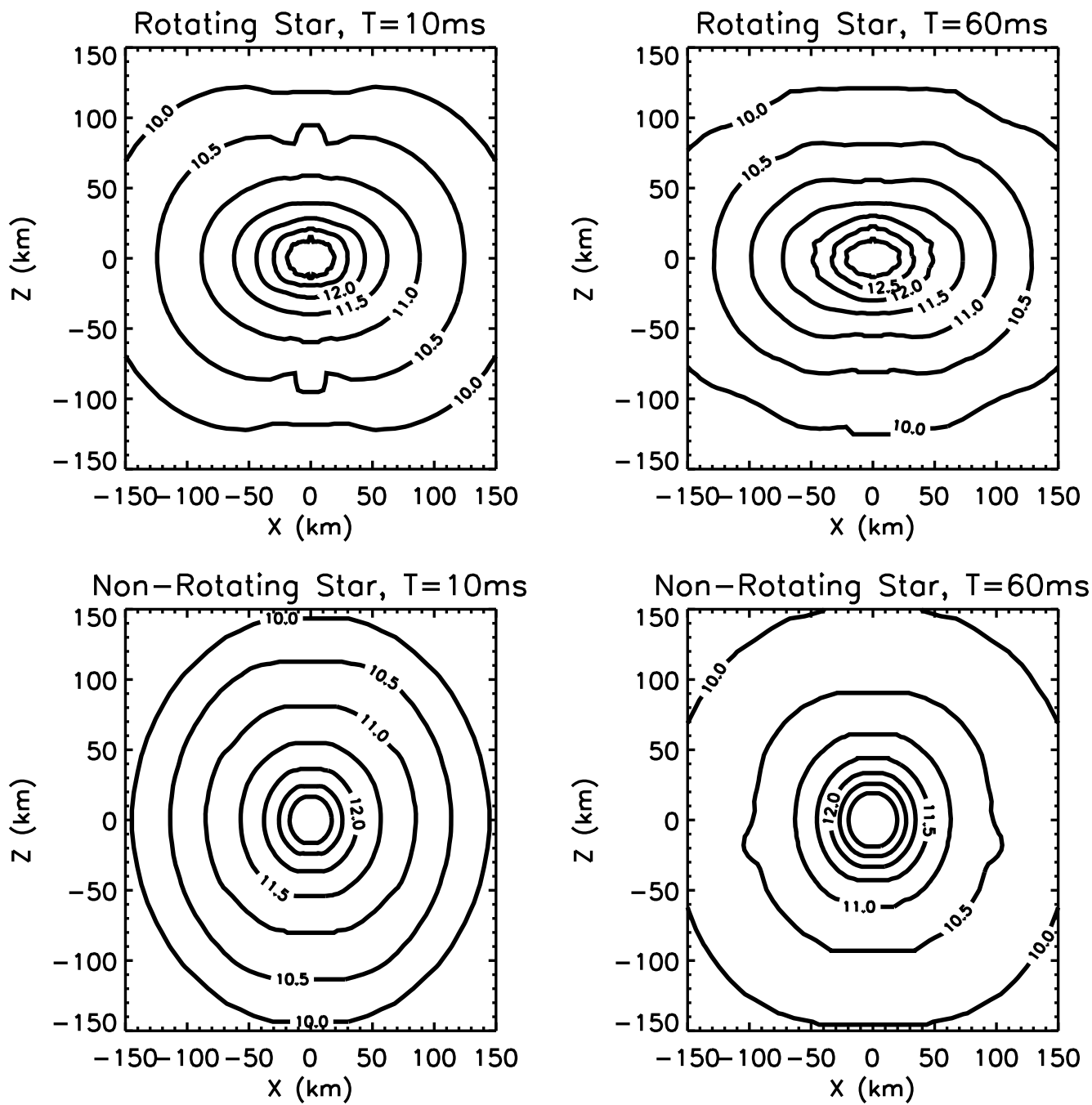}{7in}{0}{90}{90}{-370}{0}
\caption{Surfaces of constant entropy (logarithmically spaced) for 
rotating and non-rotating stars (Models 1,5) at 10 and 60\,ms after 
bounce.  60\,ms after the bounce, the axis ratio (pole/equator) of the 
proto-neutron star (densities $\gtrsim 10^{11}$\,g\,cm$^{-3}$) 
has dropped to 0.6.  At this time, the accretion shock of the rotating model 
is at $\sim 160,300$\,km, respectively, for the poles,equator.  In the 
case of a non-rotating star, the spherically symmetric accretion shock is 
at $\sim 300$\,km.}
\end{figure}
\clearpage

\begin{figure}
\plotfiddle{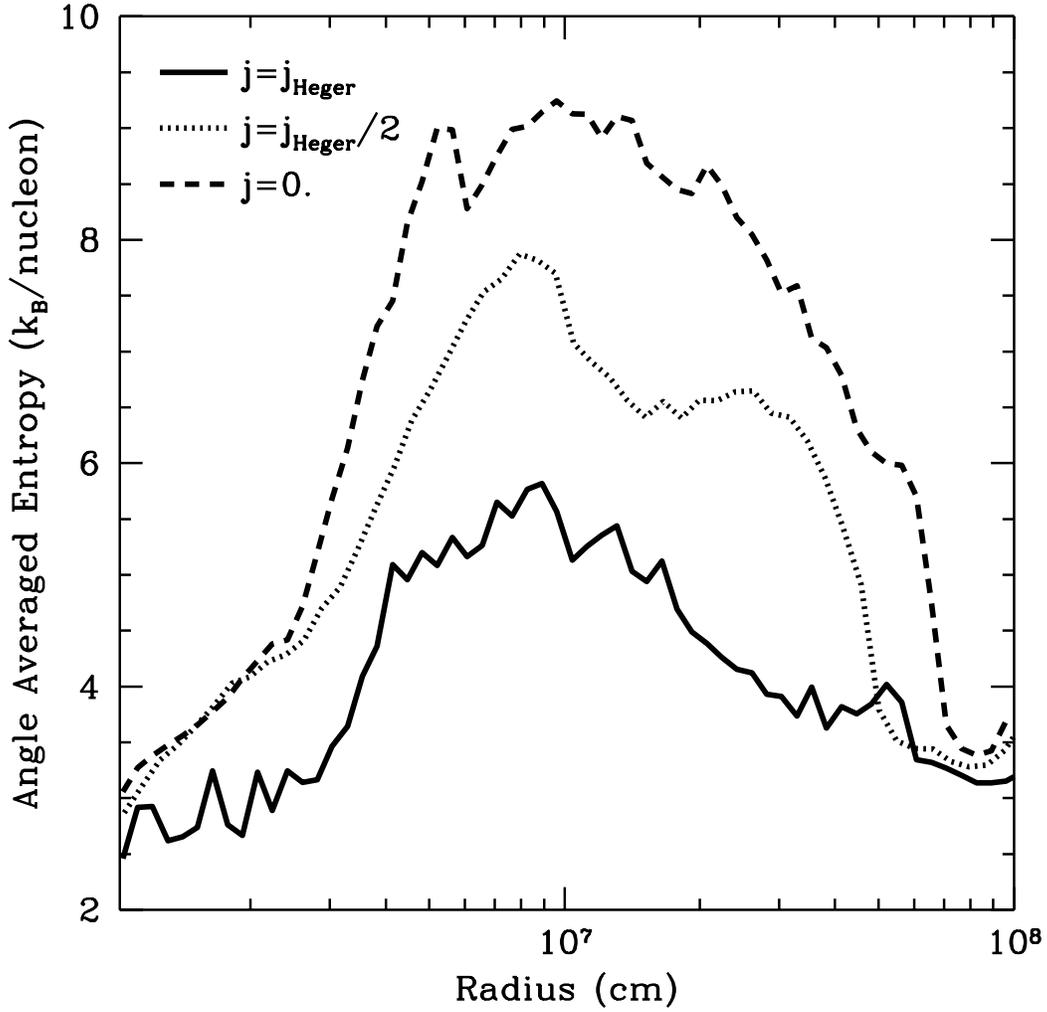}{7in}{0}{70}{70}{-215}{-10}
\caption{Angle-averaged entropy for stars with a range of initial 
angular momentum (Models 1,5,6).  As the angular momentum increases, 
the entropy profile becomes shallower.  The entropy gradient seeds 
the convection which must overcome the ram pressure of the accretion 
shock to produce a supernova explosion.  Convection in the case of 
rotating stars will be less vigorous and these stars will explode 
later.}
\end{figure}
\clearpage

\begin{figure}
\plotfiddle{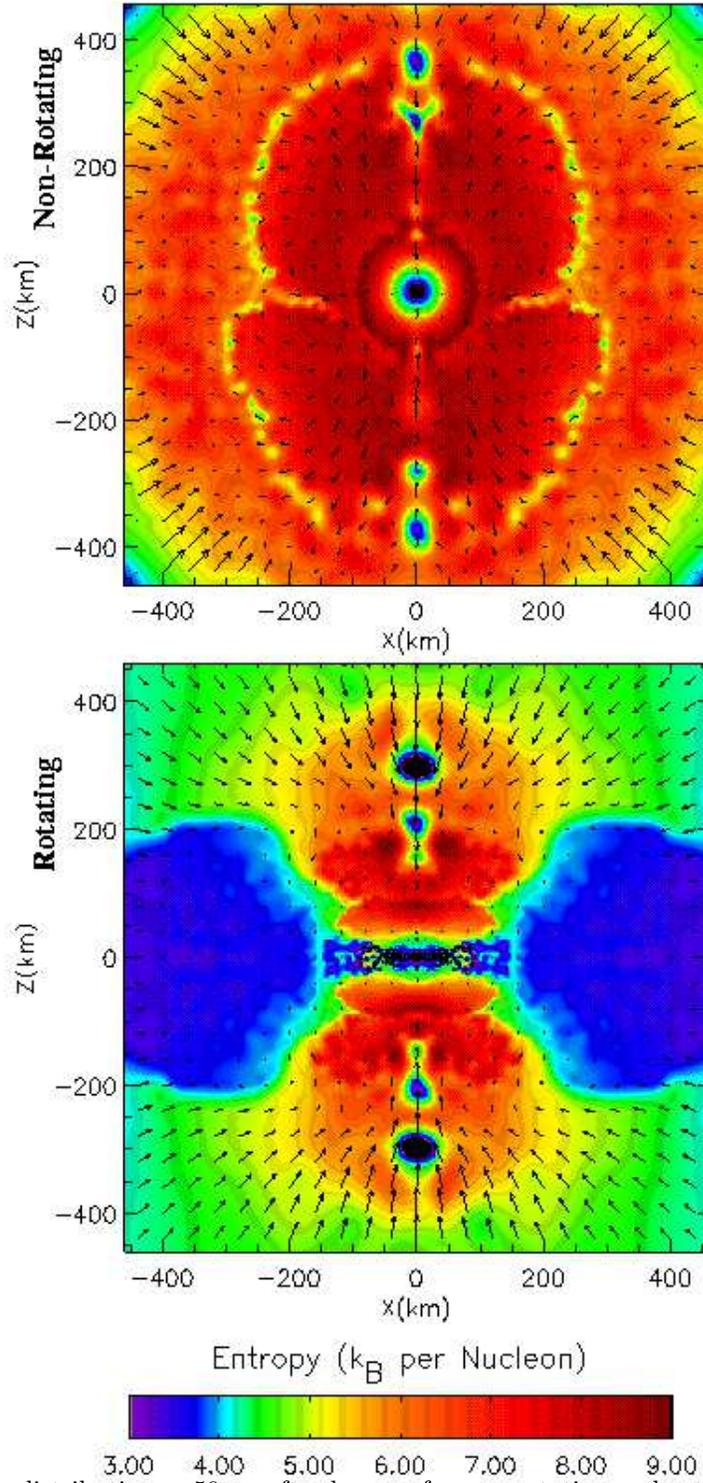}{7in}{0}{70}{70}{-190}{-10}
\caption{Entropy distribution $\sim 50$\,ms after bounce for non-rotating 
and rotating cores.  For the rotating model, the large increase in 
entropy caused by the bounce shock is limited to the polar region.  
The overlying velocity vectors show that the non-rotating model 
has already developed vigorous convection.}
\end{figure}
\clearpage

\begin{figure}
\plotfiddle{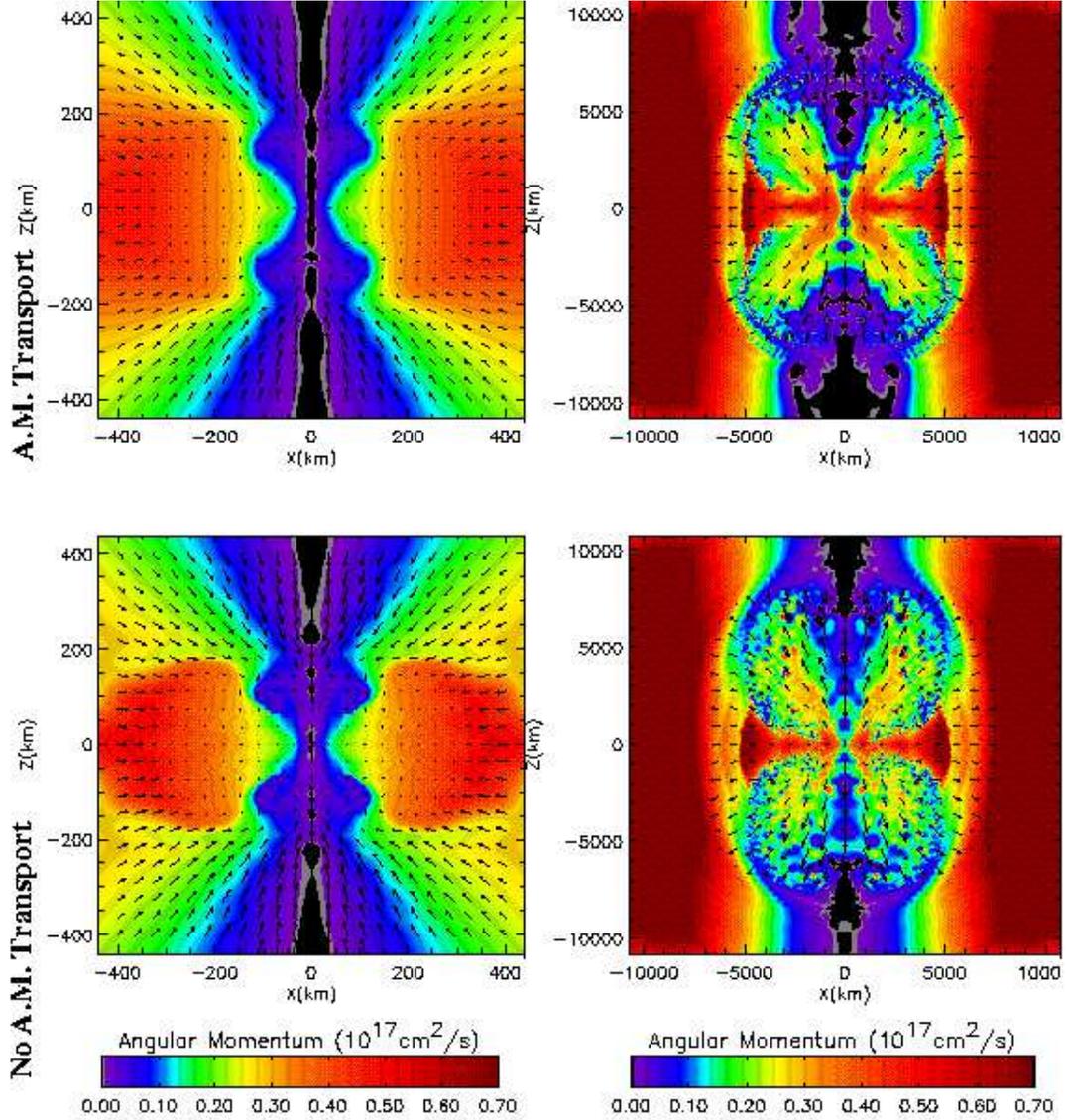}{7in}{0}{70}{70}{-230}{-30}
\caption{Angular momentum distribution roughly $\sim 50$\,ms and 
at the end of the simulation for both Models 1 and 4.  The large 
angular momentum gradient along the equator must be overcome 
to drive convection in the equatorial region.  Angular 
momentum transport in Model 4 smoothes the angular momentum 
profile, but because the rotation period is still roughly 
0.1 s by the end of the simulation, the amount of angular 
momentum in the two models is similar.  At late times, 
both models show the presence of a disk around the proto-neutron 
star.}
\end{figure}
\clearpage

\begin{figure}
\plotfiddle{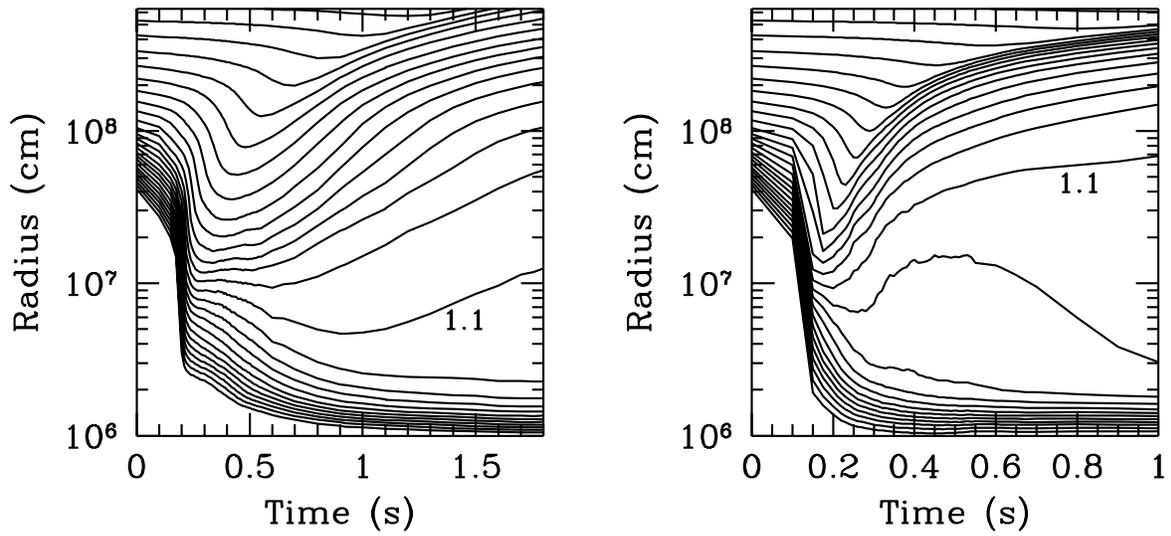}{7in}{0}{80}{80}{-240}{-200}
\caption{Mass trajectories vs. time for the rotating (left) 
and non-rotating (right) models (Models 1 and 6).  Note that 
the rotating model explodes much later than the non-rotating 
case.  The mass point which contains 1.1 M\sun is labeled and 
every line demarks a change of 0.05M\sun.}
\end{figure}

\clearpage

\begin{figure}
\plotfiddle{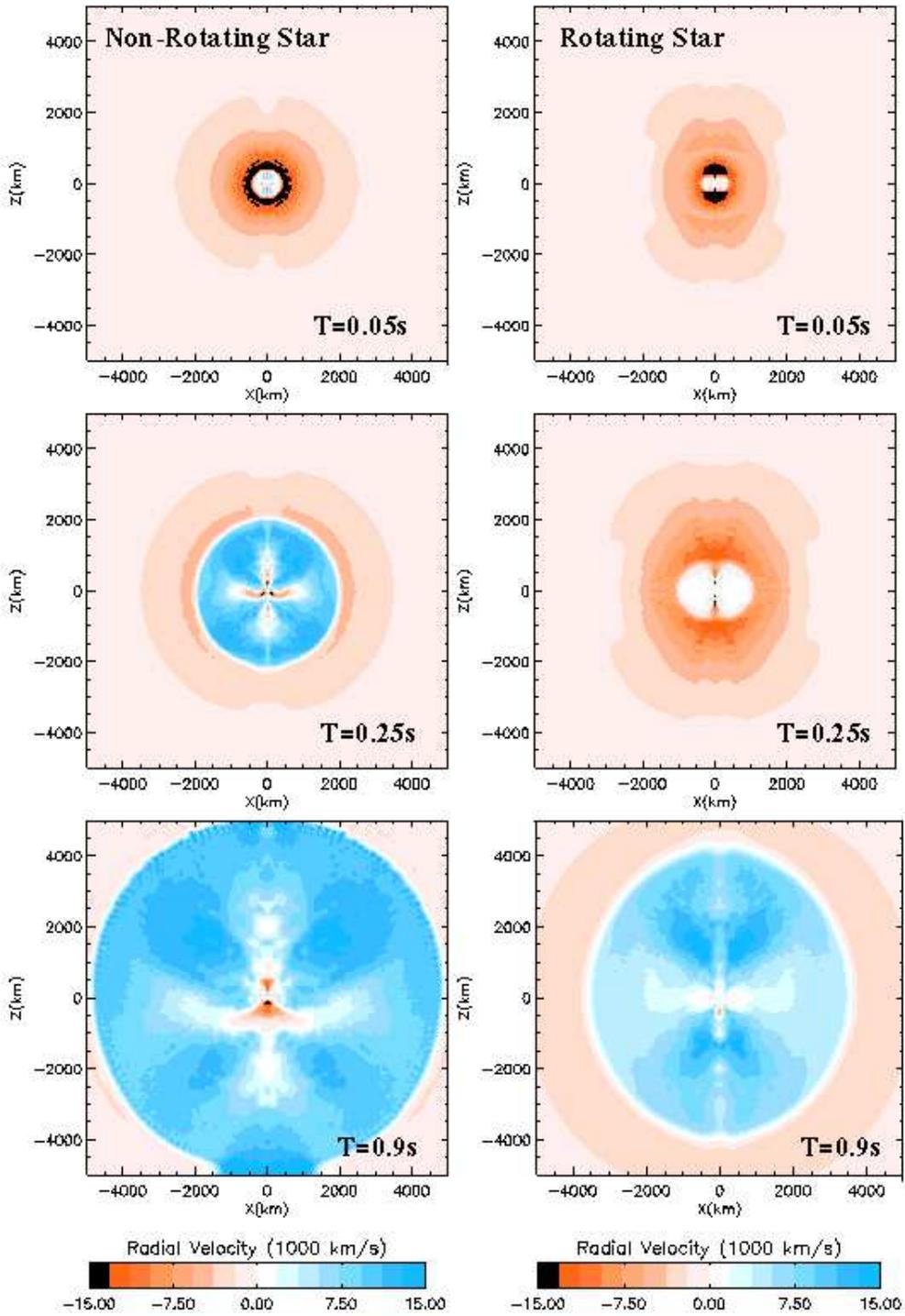}{7in}{0}{70}{70}{-220}{-10}
\caption{Radial velocity distribution of non-rotating and rotating
models 0.05, 0.25, 0.5, and 0.9\,s after bounce.  At 0.9\,s, the
non-rotating model remains essentially spherical.  The asymmetries in
the velocities are caused by the buoyant convective bubbles which are
driving the explosion.  In contrast, the rotating model already shows
strong asymmetries in the shock position and velocities.}
\end{figure}
\clearpage

\begin{figure}
\plotfiddle{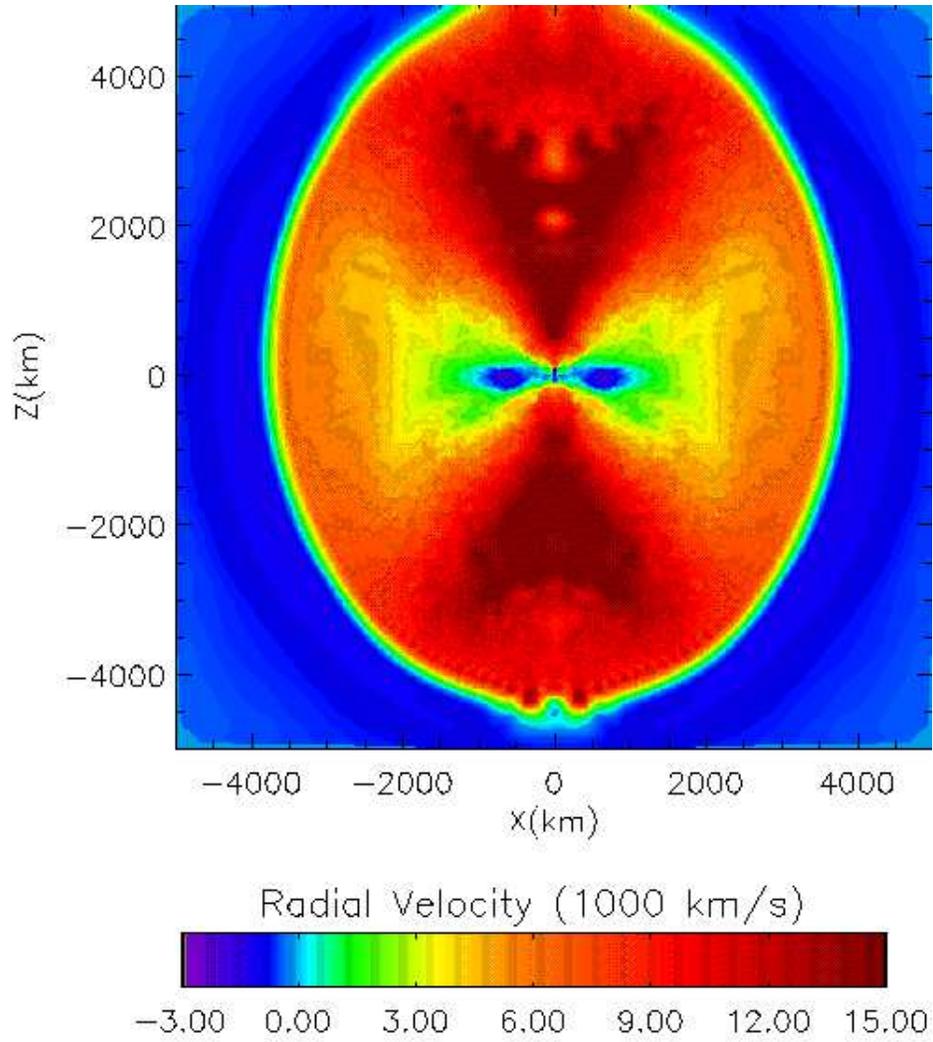}{7in}{0}{70}{70}{-245}{-30}
\caption{Radial velocity distribution of the non-rotating model (Model
1) 1.6\,s after bounce.  Note the strong jet being driven in the polar
region.  The momentum asymmetry between the upper and lower pole is
$6\times10^{39}$\,g\,cm\,s$^{-1}$, which corresponds to a neutron star
kick of roughly 30\,km\,s$^{-1}$.}
\end{figure}

\begin{figure}
\plotfiddle{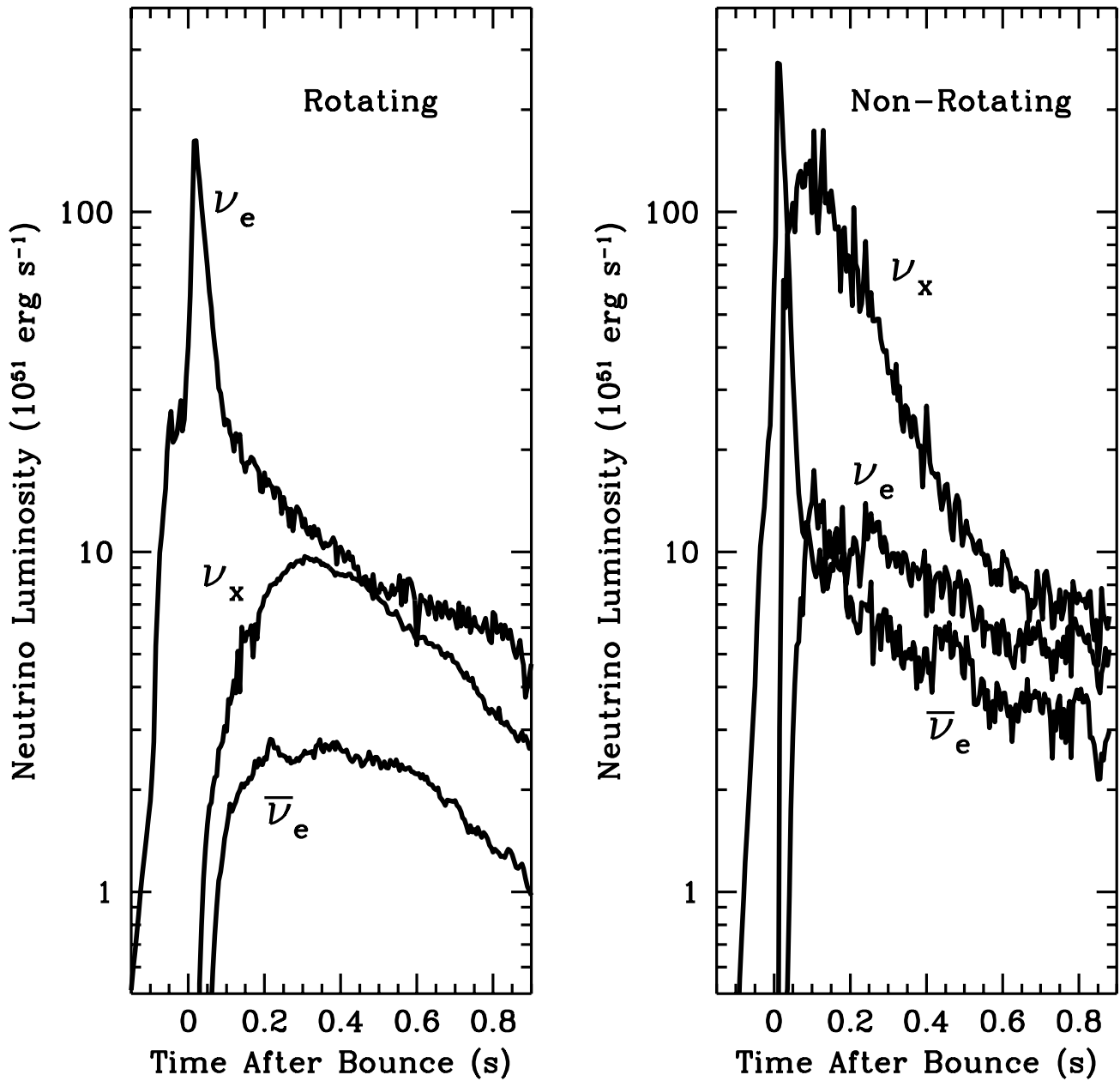}{7in}{0}{90}{90}{-280}{-100}
\caption{Neutrino luminosities versus time for a rotating and
non-rotating progenitor (models 1 and 6).  The non-rotating core has a
much larger $\mu$ and $\tau$ ($\nu_{\rm x}$) neutrino luminosity,
especially just after bounce.  This is because the non-rotating core
compresses more and, at the $\mu$ and $\tau$ neutrinosphere, the
temperature is over a factor of 1.5 higher than that of the rotating
core.  Because of the large dependence of neutrino emission on
temperature (the luminosity from pair annihilation $\propto T^9$),
this small change in temperature has large effects on the neutrino
luminosity.}
\end{figure}

\begin{figure}
\plotfiddle{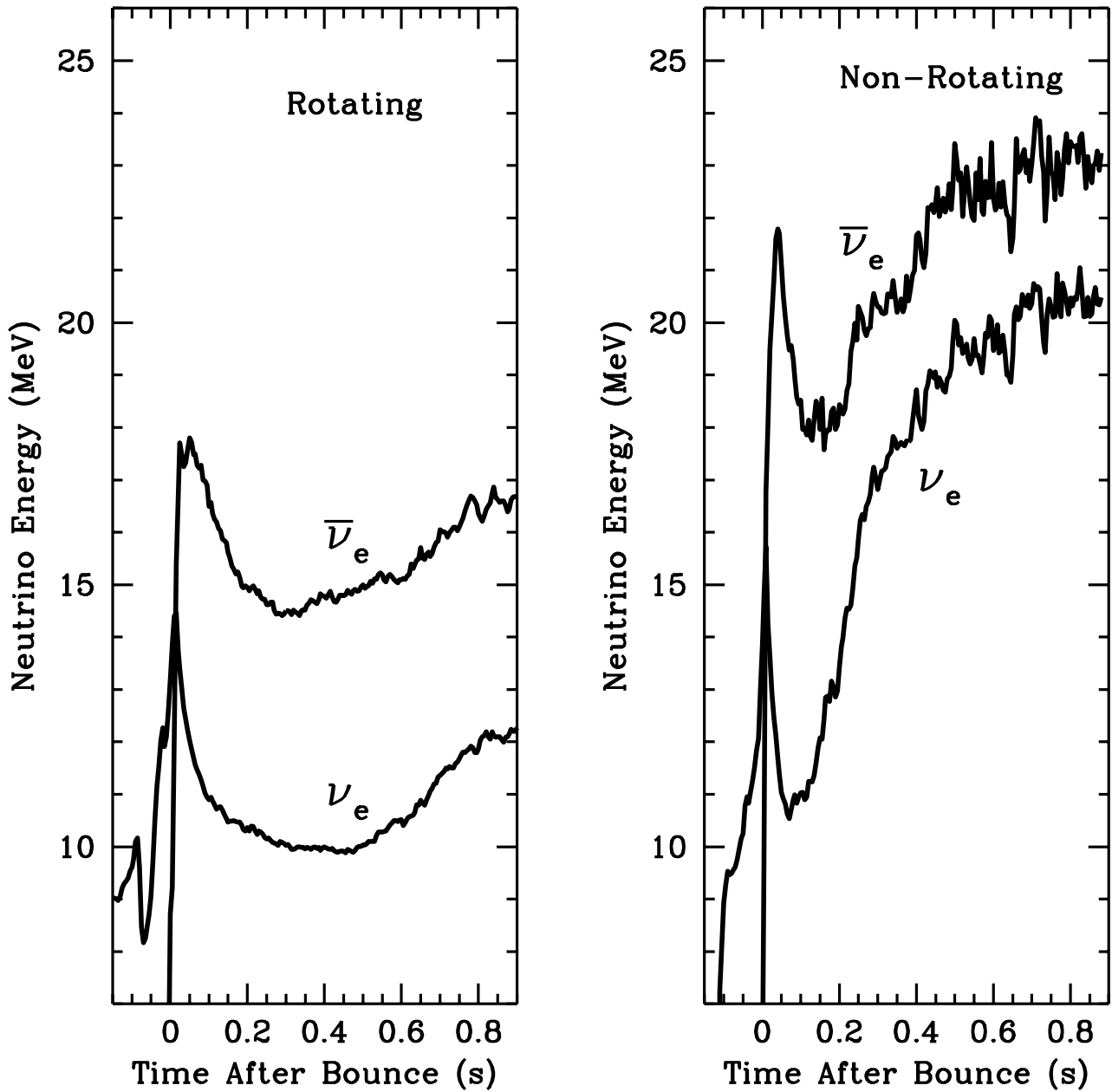}{7in}{0}{90}{90}{-280}{-100}
\caption{Electron and anti-electron neutrino energies versus time for
a rotating and non-rotating progenitor (models 1 and 6). The neutrino
energies for the non-rotating core at any given time after bounce are
much higher.  This is in part due to the larger compression, and hence
higher temperatures of the non-rotating cores.  But most of the
difference can be explained by the explosion time.  The non-rotating
star explodes $\sim 0.2$\,s after bounce, leaving a hot, bare
proto-neutron star which continues to cool by emitting neutrinos.  The
rotating core is still convecting, and its neutrinosphere is further
out where temperatures are lower.}
\end{figure}

\end{document}